\documentclass{article}
\usepackage{spconf,amsmath,graphicx,amsfonts}
\allowdisplaybreaks
\usepackage{xcolor}
\usepackage{regensky}

\usepackage[firstpage=true]{background}

\SetBgContents{\parbox{\textwidth}{\footnotesize © 2023 IEEE. Personal use of this material is permitted. Permission from IEEE must be
obtained for all other uses, in any current or future media, including
reprinting/republishing this material for advertising or promotional purposes, creating new
collective works, for resale or redistribution to servers or lists, or reuse of any copyrighted
component of this work in other works.}}
\SetBgScale{1}
\SetBgAngle{0}
\SetBgPosition{current page.north}
\SetBgVshift{-1.6cm}
\SetBgColor{black}
\SetBgOpacity{1}

\title{Improving Spherical Image Resampling through Viewport-Adaptivity}

\name{Andy Regensky, Viktoria Heimann, Ruoyu Zhang, André Kaup\thanks{This work was partly funded by the Deutsche Forschungsgemeinschaft (DFG, German Research Foundation) -- SFB 1483 -- Project-ID 442419336, EmpkinS. This work was partly funded by the Deutsche Forschungsgemeinschaft (DFG, German Research Foundation) under project number 418866191.}}
\address{Friedrich-Alexander-Universität Erlangen-Nürnberg\\Multimedia Communications and Signal Processing\\Cauerstr. 7, 91058 Erlangen, Germany}

\begin{document}
\maketitle

\begin{abstract}
    The conversion between different spherical image and video projection formats requires highly accurate resampling techniques in order to minimize the inevitable loss of information.
    Suitable resampling algorithms such as nearest neighbor, linear or cubic resampling are readily available.
    However, no generally applicable resampling technique exploits the special properties of spherical images so far.
    Thus, we propose a novel viewport-adaptive resampling (VAR) technique that takes the spherical characteristics of the underlying resampling problem into account.
    VAR can be applied to any mesh-to-mesh capable resampling algorithm and shows significant gains across all tested techniques.
    In combination with frequency-selective resampling, VAR outperforms conventional cubic resampling by more than 2 dB in terms of WS-PSNR.
    A visual inspection and the evaluation of further metrics such as PSNR and SSIM support the positive results.
\end{abstract}

\begin{keywords}
360-degree, spherical, viewport-adaptive, frequency-selective, image resampling
\end{keywords}

\vspace*{-0.5em}\section{Introduction}\label{sec:introduction}\vspace*{-0.5em}

Spherical, 360-degree or omnidirectional images and videos play a major role in the creation of immersive virtual reality experiences by allowing to present content with an omnidirectional field of view to the user.
Typically, these images and videos are captured by combining the views of multiple cameras in equirectangular projection (ERP) format to achieve an all-around field of view~\cite{Li2020, Lo2022, Cheng2022}.

However,
the ERP format comes with strong non-linear distortions especially in the areas close to the poles.
These distortions deteriorate the performance of many state-of-the-art image and video compression techniques that have been particularly designed for perspective content~\cite{Wien2017, Bross2021a}.
It has been found that other projection formats such as the cubemap projection (CMP),
the equiangular cubemap (EAC) projection, or the octahedron projection (OHP) can improve the compression efficiency significantly~\cite{Ye2020a, He2018, Xiu2020}.

\begin{figure}
  \input{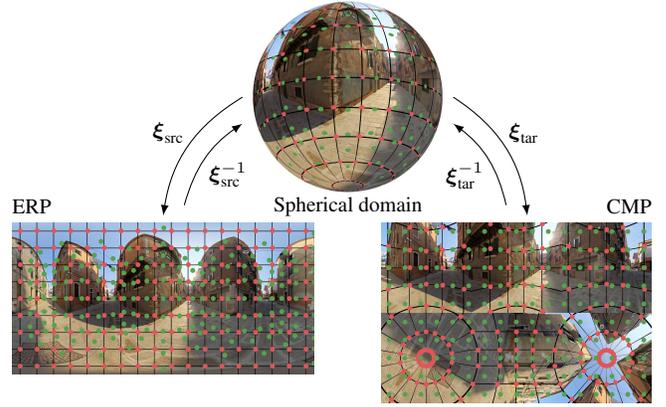}
  \vspace*{-24pt}
  \caption{Spherical image with projections to equirectangular format (left) and cubemap format (right). Dots refer to a subset of ERP (red) and CMP (green) grid coordinates. Conversion between formats requires resampling. In this context, $\boldsymbol{\xi}_\text{src}$ and $\boldsymbol{\xi}_\text{tar}$ denote the equirectangular and cubemap projection functions, respectively.}
  \label{fig:spherical_image}
\end{figure}

Fig.~\ref{fig:spherical_image} shows a spherical image in ERP and CMP projection format together with their common spherical domain representation.
The conversion between projection formats typically works by projecting all sample positions from one format to another format via their common spherical domain representation~\cite{Ye2020a}.
The resulting resampling problem from the known samples in the source projection (e.g. ERP) to the unknown samples in the target projection (e.g. CMP) can then be solved using conventional resampling techniques such as nearest neighbor, linear, cubic~\cite{Amidror2002, Alfeld1984}, or frequency-selective mesh-to-grid resampling~\cite{Heimann2020}.
However, no generally applicable resampling technique exploits the special properties of spherical images to achieve a better resampling quality so far.

To improve upon this, we propose a novel viewport-adaptive resampling (VAR) technique that builds upon the special characteristics of spherical image resampling.
By performing the resampling procedure on intermediate perspective planes,
a resampling problem is formulated that can be solved using any resampling algorithm capable of scattered data interpolation.

\section{State of the Art}\label{sec:groundwork}

When converting between projection formats, image data on a regular spaced pixel grid in source projection format $\vec{p}_\text{src} \in \mathcal{I}_\text{src}$ (e.g. ERP) needs to be resampled to a regular spaced pixel grid in target projection format $\vec{p}_\text{tar} \in \mathcal{I}_\text{tar}$ (e.g. CMP), where $\mathcal{I}_\text{src} \subset \mathbb{N}^2$ and $\mathcal{I}_\text{tar} \subset \mathbb{N}^2$ denote the set of pixel coordinates on the source grid and the target grid, respectively.

In order to resample a spherical image from one projection format to another, both the source samples and the target sample positions need to be available in a common domain.
In this context, Fig.~\ref{fig:spherical_image} shows that samples can be mapped from one projection format to the other using the corresponding projection functions.
In general, a projection function $\boldsymbol{\xi} : \mathcal{S} \rightarrow \mathbb{R}^2$ describes the relation between a point on the unit sphere and its projected position on the 2D image plane, and its inverse $\boldsymbol{\xi}^{-1} : \mathbb{R}^2 \rightarrow \mathcal{S}$ describes the inverse relationship.
The set $\mathcal{S} = \{ \vec{s} \in \mathbb{R}^3\ |\ \lVert\vec{s}\rVert_2 = 1 \}$ describes the set of all pixel coordinates on the unit sphere.

Based on this, two general possibilities exist to map the source and target coordinates to a common domain.
Either, the source coordinates are projected to the target domain yielding $\vec{p}_{\text{src}\rightarrow\text{tar}} \in \mathbb{R}^2$, or the target coordinates are projected to the source domain yielding $\vec{p}_{\text{tar}\rightarrow\text{src}} \in \mathbb{R}^2$ with
\begin{alignat}{2}
  \vec{p}_{\text{src}\rightarrow\text{tar}} &= \boldsymbol{\xi}_{\text{tar}}\left(\boldsymbol{\xi}_{\text{src}}^{-1}(\vec{p}_\text{src})\right)\quad &\forall\ \vec{p}_\text{src} &\in \mathcal{I}_\text{src},\label{eq:src_in_tar} \\
  \vec{p}_{\text{tar}\rightarrow\text{src}} &= \boldsymbol{\xi}_{\text{src}}\left(\boldsymbol{\xi}_{\text{tar}}^{-1}(\vec{p}_\text{tar})\right)\quad &\forall\ \vec{p}_\text{tar} &\in \mathcal{I}_\text{tar}.\label{eq:tar_in_src}
\end{alignat}
As the reprojected coordinates do not necessarily define a regularly spaced grid anymore, such sets of irregularly spaced pixel coordinates are denoted as \textit{mesh} in the following.
Depending on whether the resampling shall be performed in the source or the target domain, a grid-to-mesh or a mesh-to-grid resampling problem is obtained.

When performing the resampling in the source domain, the known pixel values located at $\vec{p}_{\text{src}}$ need to be resampled to the desired positions $\vec{p}_{\text{tar}\rightarrow\text{src}}$.
The resulting grid-to-mesh resampling problem can be solved using well-known algorithms such as nearest neighbor, linear and cubic resampling~\cite{Amidror2002, Alfeld1984}.

When performing the resampling in the target domain, on the other hand, the known pixel values located at $\vec{p}_{\text{src}\rightarrow\text{tar}}$ need to be resampled to the desired positions $\vec{p}_\text{tar}$.
The resulting mesh-to-grid resampling problem can again be solved using nearest neighbor, linear and cubic resampling, or using frequency-selective mesh-to-grid image resampling (FSMR), which has shown competitive results for image extrapolation, reconstruction, and resampling applications~\cite{Heimann2020, Kaup2005, Heimann2021}.

\section{Viewport-Adaptive Image Resampling}\label{sec:proposal}

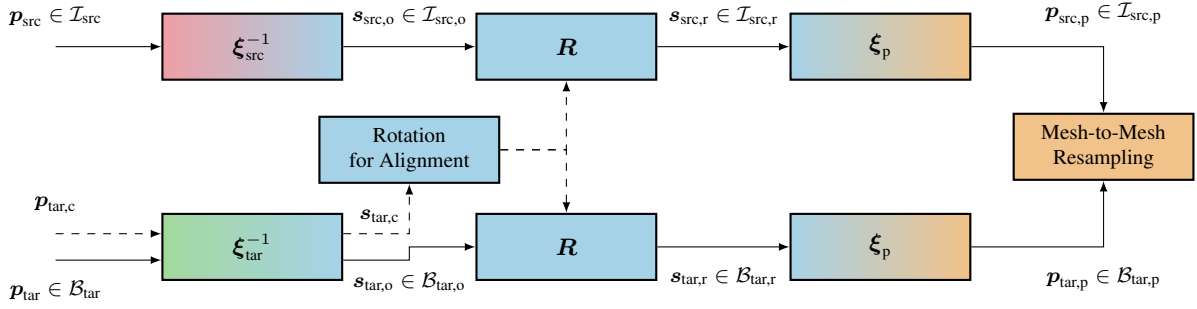
\begin{figure*}
    \centering
    \tikzstyle{block}=[draw, thick, minimum width=68pt, minimum height=25pt, align=center, node distance=50pt]
\tikzstyle{arrow}=[-latex]

\begin{tikzpicture}
  \footnotesize

  \definecolor{better_yellow}{HTML}{F6BE00}
  \definecolor{saffron}{HTML}{E6BD37}
  \definecolor{orangecrayola}{HTML}{FC7536}
  \definecolor{ruber}{HTML}{D73371}
  \definecolor{carolinablue}{HTML}{4DA5D1}
  \definecolor{lightcyan}{HTML}{D7EAE3}
  \definecolor{lightgreen}{HTML}{78ffd6}
  \definecolor{violet}{HTML}{EAAFC8}

  \definecolor{complement_red}{HTML}{EB868D}
  \definecolor{complement_orange}{HTML}{F1C289}
  \definecolor{complement_blue}{HTML}{A4BCCB}
  \definecolor{complement_green}{HTML}{7DC871}

	\colorlet{spherical_domain}{white!50!carolinablue}
  \colorlet{perspective_projection}{complement_orange}
  \colorlet{source_projection}{white!20!complement_red}
  \colorlet{target_projection}{white!30!complement_green}

  \node[block, fill=perspective_projection] (resampling) at (0, 0) {Mesh-to-Mesh\\Resampling};

  \node[block, above left = 25pt and 50pt of resampling.center, left color=spherical_domain, right color=perspective_projection] (perspective_src) {\small$\boldsymbol{\xi}_\text{p}$};
  \node[block, left = of perspective_src, fill=spherical_domain] (rotation_src) {\small$\mat{R}$};
  \node[block, left = of rotation_src, left color=source_projection, right color=spherical_domain] (projection_src) {\small$\boldsymbol{\xi}^{-1}_\text{src}$};

  \node[block, below left = 25pt and 50pt of resampling.center, left color=spherical_domain, right color=perspective_projection] (perspective_tar) {\small$\boldsymbol{\xi}_\text{p}$};
  \node[block, left = of perspective_tar, fill=spherical_domain] (rotation_tar) {\small$\mat{R}$};
  \node[block, left = of rotation_tar, left color=target_projection, right color=spherical_domain] (projection_tar) {\small$\boldsymbol{\xi}^{-1}_\text{tar}$};

  \path coordinate[below=5pt of projection_tar.west] (ctw);
  \path coordinate[above=5pt of projection_tar.west] (ctcw);
  \path coordinate[below=5pt of projection_tar.east] (cte);
  \path coordinate[above=5pt of projection_tar.east] (ctce);
  \path (rotation_tar.west) ++(0,-5pt) -- (cte) node[midway] (cte_mid) {};
  \node[block, above = 26pt of cte_mid, fill=spherical_domain] (rotation_calculation) {Rotation\\for Alignment};

  \draw[arrow] coordinate[left = 40pt of projection_src] (coords_src_start) (coords_src_start) -- (projection_src) node[pos=0] (coord_src) {};
  \draw[arrow] (projection_src) -- (rotation_src) node[midway] (sphere_src) {};
  \draw[arrow] (rotation_src) -- (perspective_src) node[midway] (rotated_sphere_src) {};
  \draw[arrow] (perspective_src.east) -| (resampling.north) node[midway] (coord_perspective_src) {};
  \draw[arrow] coordinate[left = 40pt of ctw] (coords_tar_start) (coords_tar_start) -- (ctw) node[pos=0] (coord_tar) {};
  \draw[arrow, dashed] coordinate[left = 40pt of ctcw] (coords_tar_c_start) (coords_tar_c_start) -- (ctcw) node[pos=0] (coord_c_tar) {};
  \draw[arrow] (cte) -- (cte_mid.center) |- (rotation_tar) node[midway] (sphere_tar) {};
  \draw[arrow, dashed] (ctce) -| (rotation_calculation) node[midway] (sphere_center_tar) {};
  \draw[arrow] (rotation_tar) -- (perspective_tar) node[midway] (rotated_sphere_tar) {};
  \draw[arrow] (perspective_tar.east) -| (resampling.south) node[midway] (coord_perspective_tar) {};
  \draw[arrow, dashed] (rotation_calculation.east) -| (rotation_src.south);
  \draw[arrow, dashed] (rotation_calculation.east) -| (rotation_tar.north);

  \node[above = 2pt of coord_src] {$\vec{p}_\text{src} \in \mathcal{I}_\text{src}$};
  \node[above = 2pt of sphere_src] {$\vec{s}_\text{src,o} \in \mathcal{I}_\text{src,o}$};
  \node[above = 2pt of rotated_sphere_src] {$\vec{s}_\text{src,r} \in \mathcal{I}_\text{src,r}$};
  \node[above = 2pt of coord_perspective_src] {$\vec{p}_\text{src,p} \in \mathcal{I}_\text{src,p}$};
  \node[above = 2pt of coord_c_tar] {$\vec{p}_\text{tar,c}$};
  \node[below = 2pt of coord_tar] {$\vec{p}_\text{tar} \in \mathcal{B}_\text{tar}$};
  \node[below = 4pt of sphere_tar] {$\vec{s}_\text{tar,o} \in \mathcal{B}_\text{tar,o}$};
  \node[below = 2pt of rotated_sphere_tar] {$\vec{s}_\text{tar,r} \in \mathcal{B}_\text{tar,r}$};
  \node[below = 2pt of coord_perspective_tar] {$\vec{p}_\text{tar,p} \in \mathcal{B}_\text{tar,p}$};
  \node[above left = 1pt of sphere_center_tar.center] {$\vec{s}_\text{tar,c}$};
\end{tikzpicture}
    \vspace*{-4pt}
    \caption{Schematic of the viewport-adaptive resampling procedure for one block in the target projection format. The individual processing domains are color-coded as source projection (red), target projection (green), perspective plane (orange), and spherical domain (blue).}
    \label{fig:schematic}
\end{figure*}

The proposed viewport-adaptive resampling (VAR) follows a block-based procedure in order to take the spherical nature of 360-degree images into account and to reduce the influence of the specific projection formats involved in the resampling procedure.
Each block in the target projection format is resampled on a perspective plane tangential to the block center in the spherical domain using a suitable mesh-to-mesh capable resampling technique.

In the following, sets that entail the entire range of pixels in the image in source projection format are denoted by $\mathcal{I}_\text{src}$,
and sets that entail only pixels in the regarded block of the image in target projection format are denoted by $\mathcal{B}_\text{tar}$.
Further suffixes specify the exact meaning of the sets, where the suffix \textit{o} denotes pixels on the unit sphere, the suffix \textit{r} denotes pixels on the viewport-rotated unit sphere, and the suffix \textit{p} denotes pixels on the perspective image plane.
Based on context, the sets either describe coordinates $\vec{p}$ on the 2D image plane or coordinates $\vec{s}$ on the unit sphere.

We describe the individual steps of VAR for the resampling of one block in the target image.
Fig.~\ref{fig:schematic} depicts a schematic of the described VAR procedure.
As a first step, all coordinates of the source image $\vec{p}_\text{src} \in \mathcal{I}_\text{src}$ are projected to the unit sphere using the inverse source projection $\xi_\text{src}^{-1}$, and all coordinates of the currently regarded target block $\vec{p}_\text{tar} \in \mathcal{B}_\text{tar}$ as well as the respective block center $\vec{p}_\text{tar,c}$ are projected to the unit sphere using the inverse target projection $\xi_\text{tar}^{-1}$ as
\begin{alignat}{2}
    \vec{s}_\text{src,o} &= \boldsymbol{\xi}_\text{src}^{-1}(\vec{p}_\text{src})\quad &\forall\ \vec{p}_\text{src} &\in \mathcal{I}_\text{src},\\
    \vec{s}_\text{tar,o} &= \boldsymbol{\xi}_\text{tar}^{-1}(\vec{p}_\text{tar})\quad &\forall\ \vec{p}_\text{tar} &\in \mathcal{B}_\text{tar},\\
    \vec{s}_\text{tar,c} &= \boldsymbol{\xi}_\text{tar}^{-1}(\vec{p}_\text{tar,c}).
\end{alignat}
Thus, all coordinates from the source projection format and from the target projection format are represented in one common spherical domain.

In the next step, the rotation matrix is derived such that it rotates the target block center $\vec{s}_\text{tar,c}$ to the optical center of the perspective projection.
The optical center is aligned with the $x$-axis in 3D space.
Intuitively, the orientation of the perspective projection can be understood as a virtual perspective camera positioned at the origin and oriented in the direction of the $x$-axis.
Thus, a rotation of the block center to the optical axis of the perspective projection is equivalent to rotating this virtual perspective camera to look directly at the block center.
With the spherical coordinates $(1, \theta, \varphi)$ of the block center $\vec{s}_\text{tar,c}$ on the unit sphere, the center of the regarded block can be aligned with the optical axis of the perspective projection through a rotation by $\gamma = - \varphi$ around the $z$-axis and a subsequent rotation by $\beta = \pi / 2 - \theta$ around the $y$-axis yielding the overall rotation matrix
\begin{align}
    \mat{R} &= \mat{R}_\text{y}(\beta)\mat{R}_\text{z}(\gamma) = \mat{R}_\text{y}(\pi/2 - \theta)\mat{R}_\text{z}(-\varphi),
\end{align}
where $\mat{R}_\text{y}$ and $\mat{R}_\text{z}$ denote the rotation matrix around the $y$- and $z$-axis, respectively.
Then, all source image coordinates $\vec{s}_\text{src,o} \in \mathcal{I}_\text{src,o}$ on the unit sphere and all target block coordinates $\vec{s}_\text{tar,o} \in \mathcal{B}_\text{tar,o}$ on the unit sphere are rotated by the rotation matrix $\mat{R}$ to align the coordinates that shall be resampled, with the optical axis of the perspective projection
\begin{alignat}{2}
    \vec{s}_\text{src,r} &= \mat{R}\vec{s}_\text{src,o}\quad &\forall\ \vec{s}_\text{src,o} &\in \mathcal{I}_\text{src,o},\\
    \vec{s}_\text{tar,r} &= \mat{R}\vec{s}_\text{tar,o}\quad &\forall\ \vec{s}_\text{tar,o} &\in \mathcal{B}_\text{tar,o}.
\end{alignat}
Finally, the perspective projection is applied to map the sample positions available in the source projection format and the desired sample positions in the regarded block of the target projection format to the perspective image plane
\begin{alignat}{2}
    \vec{p}_\text{src,p} &= \boldsymbol{\xi}_\text{p}(\vec{s}_\text{src,r})\quad &\forall\ \vec{s}_\text{src,r} &\in \mathcal{I}_\text{src,r},\\
    \vec{p}_\text{tar,p} &= \boldsymbol{\xi}_\text{p}(\vec{s}_\text{tar,r})\quad &\forall\ \vec{s}_\text{tar,r} &\in \mathcal{B}_\text{tar,r}.
\end{alignat}
The available samples $\vec{p}_\text{src,p} \in \mathcal{I}_\text{src,p}$ from the source projection format and the unknown samples $\vec{p}_\text{tar,p} \in \mathcal{B}_\text{tar,p}$ from the target projection format are now given on a common perspective image plane.

Resampling can then be performed using any method that is capable of mesh-to-mesh resampling.
To resample an entire source image to a different target projection format, the described procedure is repeated for all blocks in the target projection.

\vspace*{-4pt}\section{Performance Evaluation}\label{sec:performance}

To evaluate the performance of the proposed VAR technique, images given in equirectangular projection (ERP) format are resampled to cubemap projection (CMP) format and subsequently resampled back to ERP format.
This allows the assessment of the achieved resampling quality using reference-based quality metrics such as PSNR, WS-PSNR~\cite{Sun2017}, and SSIM~\cite{Wang2004}, where WS-PSNR refers to the Weighted-to-Spherically uniform PSNR that weights the error at each pixel position by the spherical area covered by that pixel position.
PSNR and WS-PSNR are calculated using the 360Lib software 360Lib-12.0~\cite{Ye2020a, 360Lib-12.0}, SSIM is calculated using the SciPy library~\cite{SciPy}.
The proposed VAR technique as well as the FSMR resampling technique are implemented in Python\footnote{The source code of our VAR and FSMR implementation is publicly available at \textit{https://github.com/fau-lms/viewport-adaptive-resampling}}.
For the evaluation of further resampling techniques capable of mesh-to-mesh resampling, we employ the well-tested nearest neighbor, linear, and cubic~\cite{Alfeld1984} interpolators from the SciPy library~\cite{SciPy}.

In a large-scale test using 590 grayscale images in ERP format from the PolyHaven HDRI asset library~\cite{PolyHaven}, we compare the resampling quality of the described mesh-to-mesh resampling techniques (\textit{nearest}, \textit{linear}, \textit{cubic} and \textit{FSMR}) with and without the use of the novel VAR technique.
The original images have a resolution of $8192 \times 4096$ pixels.
To reduce the processing time and evaluate the achievable quality for more challenging lower resolutions, the resampling procedure is performed on downscaled versions (local averaging) of the original images at resolutions of $4096 \times 2048$ and $2048 \times 1024$ pixels.
The resolution of each face in the intermediate CMP format with $3 \times 2$ face packing is calculated such that the number of samples in the ERP and CMP projection formats is approximately equal resulting in a resolution of $3456 \times 2304$ pixels for the higher ERP resolution, and a resolution of $1824 \times 1216$ pixels for the lower ERP resolution.

The blocksize of all block-based algorithms is selected such that the highest resampling quality is achieved.
This leads to a blocksize of $32 \times 32$ pixels for VAR-Nearest, VAR-Linear and VAR-Cubic, and a blocksize of $8 \times 8$ pixels for FSMR and VAR-FSMR.
The conventional nearest neighbor, linear and cubic resampling techniques are not block-based and hence are applied on the entire image.

\begin{table*}
  \centering
  \begin{adjustbox}{width=\linewidth,center}
  \small
\begin{tabular}{cl||c|c|c||c|c|c||c|c|c||c|c|c}
& & \multicolumn{3}{c||}{PSNR [dB]}& \multicolumn{3}{c||}{WS-PSNR [dB]}& \multicolumn{3}{c||}{SSIM}& \multicolumn{3}{c}{Time [s]}\\
\hline
& \makecell[r]{VAR} & OFF & ON & $\Delta$ & OFF & ON & $\Delta$ & OFF & ON & $\Delta$ & OFF & ON & $\Delta$ \\
\hline
\multirow{4}{*}{\makecell[c]{$2048\times$\\$1024$}}& Nearest & 35.38 & 35.55 & +0.17 & 37.49 & 37.77 & +0.29 & 0.9537 & 0.9553 & +0.0016 & 11 & 386 & \makecell[c]{$\times$34.94} \\
& Linear & 32.62 & 34.70 & +2.08 & 32.75 & 35.15 & +2.40 & 0.9305 & 0.9400 & +0.0095 & 182 & 745 & \makecell[c]{$\times$4.09} \\
& Cubic & 34.31 & 37.68 & +3.37 & 34.91 & 39.44 & +4.53 & 0.9579 & 0.9664 & +0.0086 & 193 & 751 & \makecell[c]{$\times$3.89} \\
& FSMR & 27.46 & 38.30 & +10.83 & 27.83 & 39.94 & +12.11 & 0.8061 & 0.9682 & +0.1621 & 9838 & 9805 & \makecell[c]{$\times$1.00} \\
\hline
\multirow{4}{*}{\makecell[c]{$4096\times$\\$2048$}}& Nearest & 35.28 & 35.41 & +0.13 & 36.48 & 36.69 & +0.21 & 0.9567 & 0.9578 & +0.0011 & 45 & 1699 & \makecell[c]{$\times$37.76} \\
& Linear & 33.50 & 34.96 & +1.46 & 33.40 & 35.00 & +1.60 & 0.9392 & 0.9446 & +0.0054 & 786 & 3159 & \makecell[c]{$\times$4.02} \\
& Cubic & 35.97 & 38.25 & +2.28 & 36.25 & 39.59 & +3.34 & 0.9683 & 0.9726 & +0.0043 & 864 & 3049 & \makecell[c]{$\times$3.99} \\
& FSMR & 27.56 & 39.37 & +11.80 & 28.59 & 40.13 & +11.54 & 0.8212 & 0.9752 & +0.1540 & 42531 & 42386 & \makecell[c]{$\times$1.00} \\
\end{tabular}

  \end{adjustbox}
  \caption{Average resampling quality and complexity after resampling 590 images from the PolyHaven test set at two resolutions from ERP format to CMP format and back to ERP format using nearest neighbor, linear, cubic and FSMR resampling techniques. Results are shown for the classical resampling procedure (VAR=OFF) and the proposed viewport-adaptive resampling procedure (VAR=ON). $\Delta$ denotes the gain of VAR with respect to classical resampling for each resampling method and quality. The quality is assessed based on PSNR, WS-PSNR and SSIM. The complexity is assessed based on the elapsed time.}
  \label{tab:var_onoff}
\end{table*}

Table~\ref{tab:var_onoff} shows the results of the described large-scale test on the PolyHaven dataset.
For each mesh-to-mesh resampling technique, the resampling qualities obtained using classical resampling procedures are denoted by VAR=OFF, the resampling qualities obtained using the proposed VAR technique are denoted by VAR=ON, and the gains achieved by VAR compared to the corresponding classical counterparts are denoted by $\Delta$.
It is clearly visible that the proposed VAR yields significant gains across all mesh-to-mesh resampling techniques, quality metrics, and resolutions.
Considering cubic resampling, VAR achieves gains of more than 3 dB in terms of WS-PSNR for both low and high resolutions.
The observed gains are especially large for the FSMR technique that shows the lowest resampling quality by a considerable margin for 360-degree projection format resampling, if applied without the proposed VAR technique.
However, if FSMR is used in conjunction with VAR (VAR-FSMR), it reaches the highest overall image quality among the field of resampling algorithms across all quality metrics.
VAR-FSMR outperforms VAR-Cubic resampling by close to 1 dB in terms of PSNR and close to 0.5 dB in terms of WS-PSNR for both resolutions.
The reported SSIM values support these observations.

\begin{figure}
  \vspace*{-10pt}
  \small
  \newlength{\framewidth}
  \setlength{\framewidth}{0.236\linewidth}
  \setlength{\tabcolsep}{2pt}
  \begin{tabular}{cccc}
    \makecell[c]{\includegraphics[width=\framewidth]{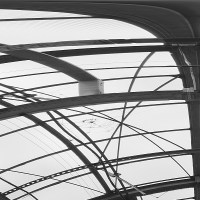}} &
    \makecell[c]{\includegraphics[width=\framewidth]{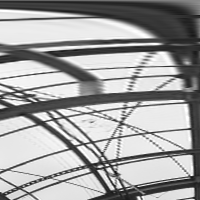}} &
    \makecell[c]{\includegraphics[width=\framewidth]{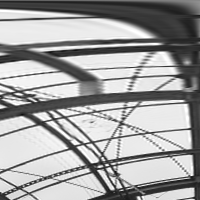}} &
    \makecell[c]{\includegraphics[width=\framewidth]{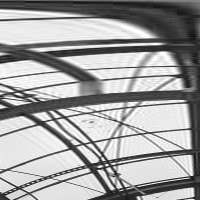}} \\
    Original & Cubic & VAR-Cubic & VAR-FSMR
  \end{tabular}
  \vspace*{-12pt}
  \caption{Crops of size $200 \times 200$ pixels from original and resampled image \textit{abandoned\_greenhouse}~\cite{PolyHaven} in ERP format using Cubic, VAR-Cubic and VAR-FSMR. Best to be viewed enlarged on a monitor.}
  \label{fig:var_crops}
\end{figure}

Furthermore, despite the considerable improvement in terms of resampling quality, the rightmost column in Table~\ref{tab:var_onoff} shows that the complexities of FSMR and VAR-FSMR, measured in terms of processing time, are almost identical.
If time is a limiting factor for the resampling procedure, the application of resampling techniques like VAR-Cubic might yield a better quality to complexity trade-off.
The application of VAR yields a roughly 4-fold complexity for linear and cubic resampling, and a roughly 35-fold complexity for nearest neighbor resampling while still processing the fastest regarding absolute elapsed time.

The large differences in relative complexity between nearest neighbor and linear/cubic resampling stem from the fact that VAR is a block-based algorithm while the conventional algorithms operate on the entire image.
For FSMR, where both the conventional FSMR and VAR-FSMR are block-based, the complexity overhead of VAR is negligible.

Fig.~\ref{fig:var_crops} shows crops from the original and resampled image \textit{abandoned\_greenhouse} using Cubic, VAR-Cubic and VAR-FSMR.
It is visible that Cubic and VAR-Cubic show staircasing artifacts on the fine structures.
These artifacts are eliminated using VAR-FSMR explaining the observed gains.

Table~\ref{tab:lib360} shows the resampling quality obtained by the proposed VAR technique and the 360Lib software, which is commonly used in standardization work in the context of video codecs.
The first frame of each sequence in the JVET 360-degree video test set~\cite{Hanhart2018} is resampled.
As visible, VAR consistently outperforms the highly optimized 360Lib software for all tested resampling algorithms with a peak gain of 1.24~dB for cubic resampling.
For comparison, the obtained quality for VAR-FSMR is shown as well.
It achieves gains of more than 2 dB compared to 360Lib's cubic resampling and 1 dB compared to VAR-Cubic resampling.

\begin{table}
  \vspace*{-10pt}
  \centering
  \small
\begin{tabular}{c||c|c|c}
& 360Lib & VAR & $\Delta$ \\
\hline
Nearest & 40.14 & 40.29 & +0.16 \\
Linear & 38.61 & 38.68 & +0.07 \\
Cubic & 43.66 & 44.90 & +1.24 \\
FSMR &  --  & 45.90 &  -- \\
\end{tabular}

  \vspace*{-4pt}
  \caption{Average resampling quality in dB based on WS-PSNR after resampling 10 images from the JVET test set~\cite{Hanhart2018} using the 360Lib and our proposed VAR for different resampling techniques. $\Delta$ denotes the gain of VAR with respect to 360Lib.}
  \label{tab:lib360}
\end{table}

\vspace*{-0.5em}\section{Conclusion}\label{sec:conclusion}\vspace*{-0.5em}

In this paper, we proposed a viewport-adaptive resampling technique that takes the spherical characteristics of 360-degree images and videos into account.
With its broad compatibility to any mesh-to-mesh capable resampling algorithm, VAR shows considerable gains across all tested algorithms.
With no increase in complexity, the combination of VAR and FSMR allows the successful application of FSMR to spherical resampling and outperforms cubic resampling by more than 2 dB in terms of WS-PSNR.

\bibliographystyle{IEEEbib}
\bibliography{ms}

\end{document}